\documentclass[pra,aps,showpacs,tightenlines,twocolumn]{revtex4}
\usepackage{epsfig}
\usepackage{amssymb}
\begin{document}
\title{DC field induced enhancement and inhibition of spontaneous emission in a cavity}
\author{G. S. Agarwal and P. K. Pathak}
\affiliation{Physical Research Laboratory, Navrangpura,
Ahmedabad-380009, India}
\date{\today}
\begin{abstract}
We demonstrate how spontaneous emission in a cavity can be
controlled by the application of a dc field. The method is
specially suitable for Rydberg atoms. We present a simple argument
based on Stark shifts for the control of emission.
\end{abstract}
 \maketitle The manipulation
of spontaneous emission has been extensively studied. Methods
involving either external fields
\cite{LANGE,LANGE2,KEITEL,GSAPRL,GSAR,ZHU,SCULLY,MOSSBERG} or
appropriate cavities \cite{FELD,HAROCHE} have been suggested. Use
of external fields enables one to control spontaneous emission via
quantum interference effects \cite{GSAR,ZHU}. Purcell
\cite{PURCELL} recognized how the emission rate in a single mode
cavity can be much higher than in free space. Several
demonstrations of the cavity enhanced spontaneous rates exists in
the literature \cite{HAROCHE,FELD}. Kleppner \cite{KLEPPNER}
discovered that the radiation rate in a cavity can be inhibited by
choosing the transition frequency such that the density of states
at this frequency is insignificant. Quantum interference between
various channels \cite{KEITEL,GSAPRL,ZHU,SCULLY} could also result
in the inhibition of emission. Further very interesting
experiments \cite{LANGE,LANGE2,ZHU,SCULLY,MOSSBERG} on the field
induced inhibition of emission in a cavity were reported. In these
experiments the applied fields were resonant with atomic
transitions. In this paper we show how a possible control of
spontaneous emission can be obtained by using dc fields. We treat
the case of atoms in a cavity and explain in rather simple terms
the origin of the control produced by the dc fields. To be precise
we are considering only the effect of dc field on the part of
decay which is due to the emission in the cavity mode. The decay
of the atom depends on the detuning between the atomic frequency
and the cavity frequency. The application of the dc field makes
the detuning dependent on the field (Stark effect) and thus the dc
field provides a control of the spontaneous emission.
 The dc field induced modification of spontaneous
emission in free space is treated in reference \cite{KEITEL}.

We next describe how to calculate the dc field induced
modification of the decay characteristics in a cavity. For our
purpose we consider a two level atom placed in a cavity  and a dc
field (or low frequency field) is injected inside the cavity . The
Hamiltonian of the system can be written as,
\begin{eqnarray}
H&=&\hbar\omega_0S^{z} +\hbar\omega_c a^{\dag}a+\hbar g \left(a
S^{+} +S^{-}a^{\dag}\right) \nonumber\\
&&+\hbar{\cal{E}}\cos{\Omega t}\left(S^{+}+S^{-}\right),
\label{ham}
\end{eqnarray}
where $\omega_0$ is atomic transition frequency, $\omega_c$ is
cavity mode frequency and $g$ is the atom cavity coupling
constant. The term ${\cal{E}}\cos{\Omega t}$ corresponds to a low
frequency field if $\Omega$ is chosen to be very small. Note that
${\cal{E}}$ has dimensions of frequency. The cavity field has been
expressed in terms of
  annihilation and creation operators $a$, $a^{\dag}$
and $S^{+}$, $S^{-}$, $S^z$ are usual atomic spin operators. We
perform master equation calculation for atom-cavity system. The
density matrix of the system $\rho$ will evolve as,
\begin{eqnarray}
\dot{\rho}=-\frac{i}{\hbar}[H,\rho]-\kappa\left(a^{\dag}a\rho-2a\rho
a^{\dag}+\rho a^{\dag}a\right), \label{mse}
\end{eqnarray}
where $2\kappa$ gives the leakage of photons. It is related to the
cavity $Q$ via $\kappa=\omega_{c}/2Q$.
 We will work in a
frame rotating with atomic frequency $\omega_{0}$
 The density
matrix in this frame is given by
\begin{equation}
\tilde{\rho}=e^{i\omega_{0}(S^{z}+a^{\dag}a)t/\hbar}\rho
e^{-i\omega_{0}(S^{z}+a^{\dag}a)t/\hbar}. \label{rho}
\end{equation}
Using Eq$(\ref{mse})$ and (\ref{rho}) we obtain the equation for
$\tilde{\rho}$
\begin{eqnarray}
\dot{\tilde{\rho}}&=&-\frac{i}{\hbar}[H_{a},\tilde{\rho}]
-\kappa\left(a^{\dag}a\tilde{\rho}-2a\tilde{\rho}
a^{\dag}+\tilde{\rho} a^{\dag}a\right)\nonumber\\
&&-\frac{i}{\hbar}[H_{d},\tilde{\rho}],
\label{mseq}\\
{\rm where}~~~~&&\nonumber\\
 H_{a}&=&-\hbar\Delta a^{\dag}a+\hbar g
\left(a S^{+}
+S^{-}a^{\dag}\right),\nonumber\\
H_{d}&=&\hbar\frac{{\cal{E}}}{2}\left\{S^{+}\left(e^{i\left(\omega_{0}+\Omega\right)t}
+e^{i\left(\omega_{0}-\Omega\right)t}\right)\right.\nonumber\\
&&\left.+S^{-} \left(e^{-i\left(\omega_{0}+\Omega\right)t}
+e^{-i\left(\omega_{0}-\Omega\right)t}\right)\right\},
\end{eqnarray}
and $\Delta=\omega_{0}-\omega_{c}$ is the detuning. We first note
that the experiments of Lange and Walther correspond to using a
microwave field and thus $\Omega\sim \omega_{0}$. The results of
Purcell and Kleppner also follow from the master equation
(\ref{mseq}). For ${\cal{E}}=0$ and $g<<\kappa$, we can derive an
equation for the atomic density matrix $\tilde{\rho}_{a}$
\begin{equation}
\tilde{\rho}_{a}=Tr_{c} \tilde{\rho}, \label{rho2}
\end{equation}
where $Tr_{c}$ is trace over cavity field, by adiabatically
eliminating cavity variables. This leads to
\begin{eqnarray}
\dot{\tilde{\rho}}_{a}=-i[\delta_{0}
S^{z},\tilde{\rho}_{a}]-\Gamma_{0}\left(S^{+}S^{-}\tilde{\rho}_{a}
-2S^{-}\tilde{\rho}_{a}S^{+}+\tilde{\rho}_{a}S^{+}S^{-}\right),
\label{old}
\end{eqnarray}
where
\begin{eqnarray}
\Gamma_{0}=\frac{g^2\kappa}{\kappa^2+\Delta^2},
~~\delta_{0}=\frac{g^2\Delta}{\kappa^2+\Delta^2}. \label{ovalue}
\end{eqnarray}
For resonant cavity $\omega_{c}=\omega_{0}$, $\delta_{0}=0$ and
the decay rate $\Gamma_{0}=g^2/\kappa$. There is cavity induced
enhancement if $g^2/\kappa$ is greater than the free space decay
rate. Note that as the cavity is detuned $(\Delta\neq0)$
$\Gamma_{0}$ decreases which is Kleppner's result for a single
mode cavity. The first experimental observation of the Purcell
effect was made by Goy {\it et al} \cite{HAROCHE}. Next we
investigate the effect of the applied dc or low frequency field.
Note that the last term in the master equation $(\ref{mseq})$ is
highly oscillating. We do time averaging for this as such terms
oscillating at the cavity frequency would not be normally
observed. The time averaging is well justified here as all other
relevant time scales $g^{-1}$, $\kappa^{-1}$, $\Delta^{-1}$ are
much larger than $(\omega_{0}\pm\Omega)^{-1}$. The inequality
$\omega_{0}>>g,~\kappa,~\Delta$ enables us to do the time
averaging in a much simpler fashion {\it i.e.} we can essentially
ignore the terms having $H_{a}$ and $\kappa$ in (\ref{mseq}). We
relegate the details of time averaging to the appendix. The
calculation leads to the following time averaged master equation
\begin{eqnarray}
\dot{\tilde{\rho}}&=& i\left[\Delta_{e}
a^{\dag}a,\tilde{\rho}\right]-ig\left[\left(a
S^{+}+S^{-}a^{\dag}\right),\tilde{\rho}\right]\nonumber\\
&&-\kappa\left(a^{\dag}a\tilde{\rho}-2a\tilde{\rho}
a^{\dag}+\tilde{\rho} a^{\dag}a\right),\label{new}\\
{\rm where}~~&&\nonumber\\
\label{stark}
&~&~\Delta_{e}=\Delta+2\omega_{0}{\cal{E}}^2/(\omega_{0}^2-\Omega^{2}).
\end{eqnarray}
 We note that
the dc field contributes to the Stark shift of the two levels in
question. We further note that these two atomic levels can also be
shifted because of the interaction of the dc field with other
levels. These can be accounted for by introducing the
polarizabilities $\alpha_e$ and $\alpha_g$ of the levels
$|e\rangle$ and $|g\rangle$ \cite{GALLAGHER,FABRE}, we can rewrite
Eq(\ref{stark}) as
\begin{equation}
\Delta_e=\Delta+\alpha_0{\cal{E}}_d^2~;~~\alpha_0=\alpha_e-\alpha_g~;
\label{newdel}
\end{equation}
where ${\cal{E}}_d$ is now the dc field in esu. The formulation of
the appendix can also be used to produce the well known
expressions for the ${\alpha}^{,s}$. The value of $\alpha_0$ is
known for many low lying as well as Rydberg transitions. The
values of $\alpha_0$ have been calculated in the literature by
converting infinite sums into the solution of differential
equations.

The Eq(\ref{new}) can be solved assuming that the atom is
initially excited and the cavity field is in vacuum state. The
Eq(\ref{new}) can be converted into a set of coupled equations in
terms of the states $|e,0\rangle$ , $|g,1\rangle$ and
$|g,0\rangle$ .
\begin{figure}[b]
\begin{center}
\includegraphics[width=3in]{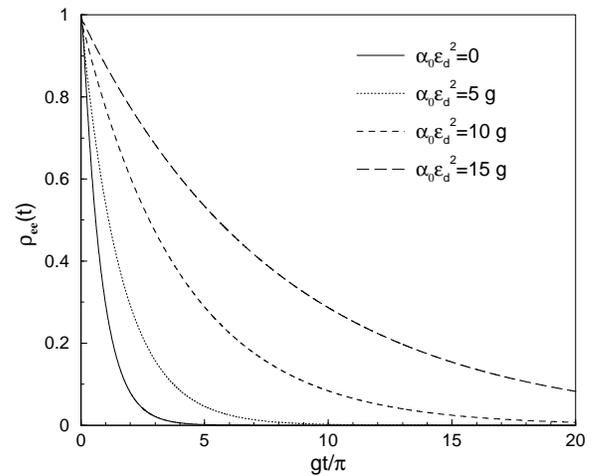}
\end{center}
\caption{The probability of the atom remaining in its excited
state, $\rho_{ee}\equiv\langle e,0|\rho|e,0\rangle$ vs time, for
$\kappa=5 g$, $\Delta=0$, $\Omega=0$, and for the different values
of the dc field ${\cal{E}}_d$.} \label{fig1}
\end{figure}
 The results of the numerical integration are shown
in the Fig\ref{fig1} for different values of the parameter
$\Delta_{e}$. Clearly there is inhibition as $\Delta_e$ increases.
The effective detuning $\Delta_e$ changes due to the applied dc
field . For a fixed cavity detuning $\Delta$ the dc field can make
$\Delta_e$ larger or smaller depending on the sign of $\Delta$.
The results can be understood by deriving analytical results in
the bad cavity limit \cite{comment} $g<<\kappa$ (and more
precisely $g^2<<\kappa^2+\Delta_e^2$). In this limit we can obtain
a simpler equation for the atomic density matrix
$\tilde{\rho}_{a}$ defined by Eq(\ref{rho2}). The final result for
the atomic system is
\begin{eqnarray}
\dot{\tilde{\rho}}_{a}=-i[\delta_{e}
S^{z},\tilde{\rho}_{a}]-\Gamma_{e}\left(S^{+}S^{-}\tilde{\rho}_{a}
-2S^{-}\tilde{\rho}_{a}S^{+}+\tilde{\rho}_{a}S^{+}S^{-}\right),
\label{master}
\end{eqnarray}
where
\begin{eqnarray}
\Gamma_{e}=\frac{g^2\kappa}{\kappa^2+\Delta_{e}^2},
~~\delta_{e}=\frac{g^2\Delta_{e}}{\kappa^2+\Delta_{e}^2}.
\label{value}
\end{eqnarray}
Here $\Gamma_{e}$ is the dc field modified decay parameter and
$\delta_{e}$ is the net frequency shift. The ratio $\eta$ of the
decays in the presence and absence of dc field is given by
\begin{eqnarray}
\eta
=\frac{\Gamma_{e}}{\Gamma_{0}}=\frac{\kappa^2+\Delta^2}{\kappa^2+\Delta_{e}^2}.
\end{eqnarray}
Clearly the dc field modifies the decay rate which depends on the
detuning. For the cavity resonant to the atomic transition
 $(\Delta=0)$, using Eq(\ref{stark}), $\eta$ reduces to
\begin{eqnarray}
\eta=\frac{\kappa^2}{\kappa^2 +\alpha_{0}^2{\cal{E}}_d^4}~~
 \approx\left(1+\frac{4{\cal{E}}^4}{\kappa^2\omega_{0}^2}\right)^{-1},~{\rm for}~\Omega=0.
 \label{ratio}
 \end{eqnarray}
It is clear from the Eq(\ref{ratio}) that dc field inhibits the
decay rate. Note that the inhibition starts becoming significant
for
\begin{eqnarray}
\alpha_0{\cal{E}}_d^{2}\sim \kappa. \label{cond}
\end{eqnarray}
 Let us estimate the condition (\ref{cond}) for $Na$ Rydberg
transition $23S_{1/2}\rightarrow 22P_{3/2}$ whose frequency is
$340GHz$ . For the sake of argument we also assume
$\alpha_0{\cal{E}}_d^{2}\sim2{\cal{E}}^{2}/\omega_0$. This
transition has a dipole moment $d\sim 10^{-15}esu$. The atom is
placed in the cavity having one mode resonant to the atomic
transition. Let us choose the cavity decay rate $\kappa=1MHz$.
 The condition (\ref{cond}) then leads to a Rabi
 frequency ${\cal{E}}$ of the order $400MHz$
which in turn requires a dc field of the order of $10^{-2}esu$. We
note that the required dc field is small enough so that the
perturbative results for the Stark shift hold. We further note
that the scalar and tensor polarizabilities are available for some
$S$ and $P$ levels of $Na$ \cite{FABRE,GALLAGHER} though the
absolute values for both $23S_{1/2}$ and $22P_{3/2}$ level are not
available in Fabre {\it et al} \cite{FABRE}. However the reported
polarizabilities for say $23P$ level are of the order of few
$MHz/(Volt/cm)^2$. Thus the condition (\ref{cond}) is realistic
and our finding that the dc field can be used to control
spontaneous emission, can be implemented by the appropriate choice
of the Rydberg transitions, [{\it cf } the condition
(\ref{cond})]. We emphasize that we are discussing the inhibition
or enhancement of spontaneous emission on a given transition which
is resonant with the cavity. This, for example, is the transition
$23S\rightarrow 22P$ in the experiments of Goy {\it et. al}
\cite{HAROCHE}. The authors of ref. \cite{HAROCHE} emphasize this
as well and it is in the spirit of the original suggestion of
Purcell \cite{PURCELL}. It must be noted that the field ionization
techniques enable one to study transitions selectively
\cite{SPENCER}.

In the case of cavities detuned from the atomic transition,
spontaneous decay is smaller and the decay rate is given by
$\Gamma= g^2\kappa/\left(\kappa^2+\Delta^2\right)$. Further
inhibition of decay rate is possible by applying dc field. When
cavity is tuned below the atomic transition frequency ($\Delta$ is
positive) then there is significant inhibition of spontaneous
decay, which increases further as the applied dc field is
increased. On the other hand when cavity is tuned above the atomic
frequency ($\Delta$ is negative) there is enhancement in the
atomic decay {\it i.e.} on increasing the value of applied dc
field the atom decays faster.
\begin{figure}[h]
\begin{center}
\includegraphics[width=3in]{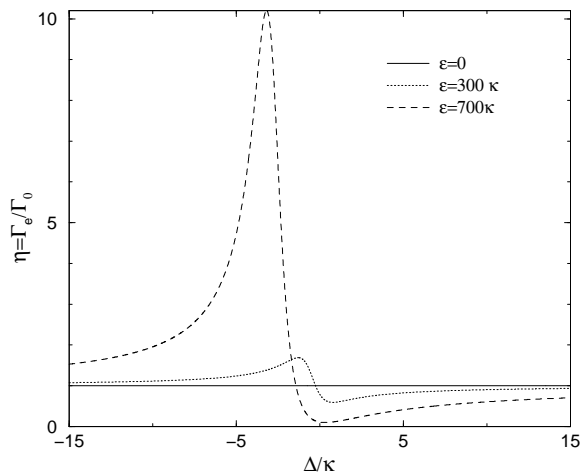}
\end{center}
\caption{The ratio ($\eta$) of the decays in the presence and the
absence of dc field vs $\Delta/\kappa$ . The parameters are
$\omega_0=3.4\times 10^5\kappa$ and $\Omega=0$.} \label{fig2}
\end{figure}
In Fig\ref{fig2} we show the behavior of the factor $\eta$ as a
function of $\Delta$ for different values of the dc field. The
enhancement as well as inhibition of spontaneous decay occurs
depending on whether the cavity is tuned above or below the atomic
frequency. The results shown in the Fig\ref{fig2} are consistent
with the results obtained by direct solution of the Eq(\ref{new}).

In conclusion we find that in presence of dc field spontaneous
emission can be inhibited significantly in the case of cavities
resonant to atomic transition. In the case of cavities having
negligible mode density around atomic frequency spontaneous
emission itself is smaller and the presence of dc field shows
significant inhibition or enhancement depending on cavity is tuned
below the atomic transition frequency or above the transition
frequency.

One of us (GSA) thanks G. Rempe and H. Walther for discussions on
this subject.
\appendix
\section{}
We outline how the time averaging is to be done. Let us consider
schrodinger equation
\begin{equation}
\frac{\partial}{\partial
t}|\psi(t)\rangle=-\frac{i}{\hbar}V(t)|\psi(t)\rangle, \label{sch}
\end{equation}
where $V(t)$ consists of rapidly oscillating terms only so that
the time average of $V(t)$ is zero. Let $|\psi\rangle$ be written
as
\begin{equation}
|\psi\rangle=|\bar{\psi}\rangle+|\phi\rangle, \label{psi}
\end{equation}
where $|\bar{\psi}\rangle$ is time averaged part and
$|\phi\rangle$ is the rapidly oscillating part. On substituting
(\ref{psi}) in Eq (\ref{sch}) we find that to the lowest order in
$V(t)$
\begin{eqnarray}
&&|\phi\rangle=-\frac{i}{\hbar}\int_{0}^{t}V(\tau)d\tau
|\bar{\psi}\rangle,\\
{\rm and}~~&&\nonumber\\ &&\frac{\partial}{\partial
t}|\bar{\psi}(t)\rangle=-\frac{i}{\hbar}\bar{V}(t)|\bar{\psi}\rangle,\\
{\rm where}~~&&\nonumber\\
&&\bar{V}(t)=-\frac{i}{\hbar}\overline{V(t)\int_{0}^{t}V(\tau)d\tau}.
\label{app}
\end{eqnarray} The field induced shift term in (\ref{new}) is
obtained by using Eq(\ref{app}).


\begin{thebibliography}{99}
\bibitem{comment} Purcell's formula assumes that bad cavity is
such that cavity mode remains well defined.
\end{thebibliography}
\end{document}